\documentstyle[onecolumn]{mn}
\input{epsf.tex}
\newcommand{\mincir}{\raise -2.truept\hbox{\rlap{\hbox{$\sim$}}\raise5.truept
\hbox{$<$}\ }}
\newcommand{\magcir}{\raise -2.truept\hbox{\rlap{\hbox{$\sim$}}\raise5.truept
\hbox{$>$}\ }}
\newcommand{\minmag}{\raise-2.truept\hbox{\rlap{\hbox{$<$}}\raise 6.truept\hbox
{$>$}\ }}
\newcommand{\be}{\begin{equation}}
\newcommand{\ee}{\end{equation}}
\newcommand{\ba}{\begin{eqnarray}}
\newcommand{\ea}{\end{eqnarray}}
\newcommand{\brr}{\begin{array}}
\newcommand{\err}{\end{array}}
\newcommand{\bc}{\begin{center}}
\newcommand{\ec}{\end{center}}

\newcommand{\hm}{\,h^{-1}{\rm Mpc}}

\title{Evolution of the two-point correlation function in the Zel'dovich
approximation}
\author[Porciani]
{Cristiano Porciani\footnotemark \\
SISSA, Scuola Internazionale di Studi Superiori Avanzati,
via Beirut 2-4, I--34014 Trieste, Italy}
\begin{document}

\maketitle

\begin{abstract}
We study the evolution of the mass autocorrelation function 
by describing the growth of density fluctuations through
the Zel'dovich approximation.
The results are directly compared with the predictions of the scaling
hypothesis for clustering evolution extracted from numerical simulations
(Hamilton et al. 1991),
as implemented by Jain, Mo \& White (1995).
We find very good agreement between the correlations
on mildly non--linear scales and on completely linear scales.
In between these regimes, we note that the
density fields evolved through the Zel'dovich approximation 
show more non--linear features than predicted by
the scaling ansatz which is, however, forced to match the linear evolution
on scales larger than the simulation box.
In any case, the scaling ansatz by Baugh \& Gazta\~naga (1996), calibrated
against large box simulations agrees better with ZA predictions
on large scales, keeping good accuracy also on intermediate scales.

We show that 
mode--coupling is able to move the first zero crossing of $\xi(r)$ as time
goes on. A detailed fit of the time dependence of this shifting is given for a
CDM model.
The evolution of the cross correlation of the density fluctuation field
evaluated at two different times is also studied. The possible
implications of the
results for the analysis of the observed correlation function
of high redshift galaxies are discussed.  
\end{abstract}

\begin{keywords}
galaxies: clustering --
cosmology: theory -- large-scale structure of Universe 
\end{keywords}

\section{Introduction}

\footnotetext {e--mail: porciani@sissa.it}

In the last two decades, redshift surveys provided a wealth of informations
about the spatial distribution of local galaxies,
revealing the existence of  large--scale structures.
The most widely used statistical tool to quantify
the degree of clustering has been the galaxy two--point correlation function,
both in its angular $w_g(\theta)$ and spatial
$\xi_g(r)$ versions 
(see, e.g., Peebles 1980).
As previous studies were confined to the nearby universe, nowadays
new observational resources permit to extend the correlation 
analysis to deeper samples.
In fact,
the Canada--France Redshift Survey has recently provided the new
opportunity 
to investigate the clustering properties of galaxies out to redshifts 
$z \sim 1$ (Le F\`evre et al. 1996).
Moreover, only lately
 it has been possible to analyse the angular distribution
of faint galaxies by using
the Hawaii Keck K--band survey (Cowie et al. 1996) and the Hubble Deep
Field data (Villumsen, Freudling \& da Costa 1997). 
Therefore, we are now confident of 
detecting a direct signature of redshift dependence in the 
observed correlation function.
For this reason, the theoretical analysis of the
evolution of the mass two--point correlation function, $\xi(r)$,
is becoming a fundamental topic of modern cosmology.
However,
it is worth stressing that
 the interpretation of the observational data is not
 immediate:
 before
obtaining the ``real'' change of the large--scale structure one
has to consider the possible evolution of the galaxy population 
(as well as the related selection effects)
and of the
bias factor that formally relates $\xi_g$ to $\xi$
(see, e.g., Matarrese et al. 1997).

The observational results should then be compared
to the predictions of the existing models for structure 
formation. One of the several issues involved in this comparison
is represented by the lack of a standard description of clustering
evolution: analytical
treatments are generally unable to manage this fully non--linear problem
 while numerical simulations are limited in resolution.
However, new light has been recently shed on this argument.
Hamilton et al. (1991) suggested that the correlation function obtained
through N-body simulations of an Einstein--de Sitter universe,
in which the structure develops hierarchically, 
can be easily reproduced by applying a
non--local and non--linear transformation to the linear $\xi(r)$.
This ansatz has been refined and extended
to more general cosmological scenarios by a number of authors
(Peacock \& Dodds 1994, Jain, Mo \& White 1995, Peacock \& Dodds 1996).
Moreover it is possible to give theoretical arguments that account
for the scaling hypothesis (Nityananda \& Padmanabhan 1994).

The main purpose of this paper is to compare the predictions of the Zel'dovich
approximation (Zel'dovich 1970) with the scaling ansatz
formulated in the
version of
Jain, Mo \& White (1995, hereafter JMW). 
Actually, it would be very interesting to 
obtain all the details of the semi--empirical
scaling relationship 
in the framework of the gravitational
instability scenario. However, in the absence of a model for the
advanced phases
of clustering evolution, we are forced to analyse only
the onset of non--linear dynamics.

From the theoretical point of view, the evolution  of the two--point
correlation function is strictly related to the dynamical development
of the density field
$\varrho({\bf x},t)$. When the dimensionless density contrast
$ \delta ({\bf x},t)=[\varrho({\bf x},t)-\bar \varrho(t)]/\bar \varrho(t)$
is much smaller than unity, the growth of the fluctuations can be 
followed performing a perturbative approach (see, e.g., Peebles 1980, 
Fry 1984, Scoccimarro \& Frieman 1996a). At the lowest order (linear theory),
the different Fourier modes
of $\delta({\bf x},t)$ evolve independently provided
that their power spectrum is less steep than
$k^4$ at small $k$ (Zel'dovich 1965, Peebles 1974).
As the fluctuations grow, however, the interactions
between different modes become more and more important.
The effect of this mode--coupling on the two--point statistics has been
studied by many authors using higher--order than linear terms 
in the perturbation expansion.
Juskiewicz, Sonoda \& Barrow (1984) computed the second--order contribution to
$\xi(r)$ for an exponentially smoothed linear spectrum $P(k)\propto k^2$,
finding that non--linear interactions among long wavelength modes
act as a source for short $\lambda$ perturbations.    
As a matter of fact, they found a substantial decrease of the characteristic
scale of clustering with the evolution.
However Suto \& Sasaki (1991) and Makino, Sasaki \& Suto (1992), analysing
exponentially filtered scale--free spectra, found that second--order effects
can either suppress or enhance the growth of perturbations on large scales,
depending on the shape and the amplitude of the fluctuation spectrum.
In particular Makino, Sasaki \& Suto (1992), modelling a CDM spectrum with
two different power laws, concluded that the effects of
mode--coupling are generally very small and completely negligible on
scales $r \magcir 20 \hm$ 
(where $h$ denotes the Hubble constant in units of $100
\,{\rm km\,s^{-1}\,Mpc^{-1}}$).
The second--order correction to the ``true'' linear CDM spectrum
has been calculated
by Coles (1990) who computed also the  respective correlation function.
The results show that, for moderate evolution,
the large--scale distortions are of no importance,
while later (for $\sigma_8 \magcir 1$,
where $\sigma_8$ represents the {\em rms} linear mass fluctuation in spheres
of radius $8 \hm$)
non--linear effects can increase the clustering strength on scales
$r > 35 \hm$;
for example, the first zero crossing of $\xi(r)$ can be significantly
shifted with
respect to linear predictions.
Similar results were obtained by Baugh \& Efstathiou (1994) who also found 
good agreement with the output of numerical simulations.
However,
Jain \& Bertschinger (1994)
pointed out that the perturbative approach
is able to reproduce the N--body outcomes only at early times
($\sigma_8 \mincir 0.5-1$). 
Moreover, the recent analysis
applied to scale--free spectra by  Scoccimarro \& Friemann
(1996b) showed
that the validity of perturbation theory 
is restricted to a small range of spectral indices.

In this paper, we want to study the 
non--linear evolution 
of the mass autocorrelation function by describing the growth of
density fluctuations through the Zel'dovich approximation
(hereafter ZA). 
In effect, Eulerian second--order perturbation theory may break down once 
the mass variance becomes sufficiently large. On the other hand, we know that
ZA, especially in its ``truncated'' form, 
is able to reproduce fairly well the outcomes of
N--body simulations even in the mildly non--linear regime 
(Melott, Pellman \& Shandarin 1994).
The main advantage of ZA over other dynamical approximations
(for a recent review see, e.g., Sahni \& Coles 1995)
is that it permits analytical investigations ensuring at the same time
good accuracy, at least for quasi--linear scales.
The pioneering analysis by Bond \& Couchmann (1988) showed
that ZA is able to predict
 the shifting of the first zero crossing of the correlation function.
In Section 3 we will give a detailed
quantitative description of this effect. 
Other features of the mass two--point correlation function in ZA have
been discussed by Mann, Heavens \& Peacock (1993, hereafter MHP).
Moreover,
the related evolution of the power spectrum has been studied by
Taylor (1993), 
Schneider \& Bartelmann (1995) and Taylor \& Hamilton (1996).
These authors
showed that ZA is able to describe the generation of small--scale
power through mode coupling, at least at early times.
Besides Fisher \& Nusser (1996) and Taylor \& Hamilton (1996)
succeeded in computing
the power spectrum
also in redshift space.

This paper is organized as follows. In Section 2 we briefly introduce the 
Zel'dovich approximation while 
in Section 3 we compute 
the cross  correlation function between the mass density field
evaluated at two different times. The usual two--point correlation function
is obtained as a particular case of this more general quantity. 
The redshift evolution of $\xi(r)$ in a
CDM model is the last subject of Section 3.
In Section 4 we compare the predictions of ZA with the
scaling ansatz of JMW.
In Section 5 we use our results to evaluate the correlation
function of a collection of objects sampled by an observer in a
wide redshift interval of his past light cone. 
We then propose a simplified scheme to compute this quantity so as
to improve another approximation presented in the literature.
A brief summary is given in Section 6.

\section{The Zel'dovich approximation}

Let us consider a set of collisionless, self--gravitating particles
in an expanding universe with scale factor $a(t)$.
We can describe the motion of each point--like particle
writing its actual (Eulerian) comoving position, ${\bf x}$, at time $t$
as the sum of
its initial (Lagrangian) comoving position, ${\bf q}$,
plus a displacement:
\be
{\bf x}({\bf q},t)={\bf q}+{\bf S}({\bf q},t).
\label{Eq:lagrange}
\ee 
The displacement vector field ${\bf S}({\bf q},t)$ represents
the effect of density perturbations on the trajectories.
The Zel'dovich approximation is obtained by assuming the separability
of the temporal and spatial parts of ${\bf S}({\bf q},t)$
and by requiring equation (\ref{Eq:lagrange}) to give
the correct evolution of $\delta({\bf x},t)$
in the linear
regime.
Considering only the growing mode for
a pressureless fluid, one gets
(Zel'dovich 1970):
\be
{\bf S}({\bf q},t)= - b(t) {\bf \nabla}
\phi \big|_{\mathbf q}
\label {Eq:zeld}
\ee
where
$b(t)$ is the linear growth factor and
$\phi({\bf q})$ represents the initial peculiar velocity potential
that at the linear stage is proportional to the gravitational potential
$\Phi_0({\bf q})$.
The Zel'dovich approximation can be also extracted from a fully Lagrangian
approach to the evolution of density fluctuations
(Buchert 1989, Moutarde et al. 1991, Bouchet et al. 1992, Buchert 1993,
 Catelan 1995).
In this case, ZA
 corresponds to the first order solution provided that 
the initial velocity field is irrotational and the
initial peculiar velocity 
and acceleration fields are everywhere parallel.

Equations (\ref{Eq:lagrange}) and (\ref{Eq:zeld}) define a mapping from
 Lagrangian
to Eulerian space that develops caustics as time goes on
(Shandarin \& Zel'dovich 1989). However, the
``Zel'dovich fluid'' is a system with infinite memory: even after the
intersection of two trajectories, the motion of the particles is determined
by their initial conditions according to equation 
(\ref{Eq:zeld}).
The lack of self--gravity between intersecting streams causes the
forming structure
to be rapidly washed out. This is a severe problem especially in hierarchical
models of structure formation,
where caustics appear early on small scales causing ZA
to become soon inaccurate.
Nevertheless Coles, Melott \& Shandarin (1993)
showed that a modified version of ZA, the ``truncated'' ZA, obtained
by smoothing the initial conditions, is able to reproduce with good accuracy
the density distributions obtained from numerical simulations.
Melott, Pellman \& Shandarin (1994) found that the optimal
version of the truncation procedure is accomplished by using a Gaussian window
to smooth the linearly extrapolated power spectrum of the
density fluctuation field $b^2(t) P(k)$:
\footnote{We set $b=1$ at
the present epoch.}
\be
P_{T}(k,t)=b^2(t) P(k)\exp{\left[ -k^2 R_f^2(t)\right]}
\label{Eq:trunc}
\ee
where the filtering radius $R_f(t)$ increases with time being
 related to the 
typical scale going non--linear.
The success of this approximation can be justified by noticing that
the non--linearly evolved gravitational potential resembles
its smoothed linear counterpart 
(Pauls \& Melott 1995).
In the following we will adopt the filtering prescription given in equation
(\ref{Eq:trunc}).
 
\section{The two-point correlation function in the Zel'dovich approximation}

Assuming that initially the mass is evenly distributed in
Lagrangian space, implies that the Eulerian density field
is related to the Lagrangian displacement field via the relation:
\be \varrho({\bf x},t)= \bar \varrho(t) \int 
 d^3q \, \delta _D \left[{\bf x}-{\bf q}-{\bf S}({\bf q},t)\right],
\label {Eq:rho-S}
\ee 
where $ \delta_D({\bf x})$ denotes the three--dimensional Dirac delta function.
For purposes that will be clarified in Section 5, we are
interested in computing the cross correlation function between the
density contrast field evaluated at two different times:
\be
 \langle \delta({\mathbf x}_1,t_1) \delta({\mathbf x}_2,t_2) \rangle =
\langle  \int d^3q_1 d^3q_2 
\, \, \delta_D \left[{\mathbf x}_1-{\mathbf q}_1-{\mathbf S}({\mathbf q}_1,
t_1)\right]
 \delta_D \left[{\mathbf x}_2-{\mathbf q}_2-{\mathbf S}({\mathbf q}_2,t_2)
\right]
\rangle -1
\label{Eq:deldel}
\ee
where $\langle \cdot \rangle$ represents the average over an ensemble of 
realizations.
Before going any further,
it is convenient to
Fourier transform the Dirac delta functions in equation (\ref{Eq:deldel})
obtaining:
\be
1+\langle \delta({\mathbf x}_1, t_1) \delta({\mathbf x}_2,t_2) \rangle=
 \int d^3q_1 d^3q_2 \, \, {d^3 w_1 \over (2 \pi)^3}\, {d^3 w_2
\over (2 \pi)^3} \exp \left[ i \sum _{j=1}^2 {{\mathbf w}_j \cdot 
({\mathbf x}_j-{\mathbf q}_j)} \right]
\langle \exp \left[ -i \sum_{\ell=1}^2 {\mathbf w}_\ell\cdot {\mathbf S}
({\mathbf q}_\ell,t_\ell) 
\right] \rangle.
\label{Eq:deldel2}
\ee
We then use equation (\ref{Eq:zeld}) to introduce ZA
into equation (\ref{Eq:deldel2}).
In such a way, by assuming, as usual, that $\phi({\bf q})$ is
a statistically homogeneous and isotropic Gaussian field,
uniquely specified by its power spectrum $P_\phi(k)\propto P(k)/k^4$,
the ensemble average contained in equation
(\ref{Eq:deldel2})
can be written as a functional integral:
\be
\langle \exp \left[ -i \sum_{\ell=1}^2 {\mathbf w}_\ell\cdot {\bf S}
({\bf q}_\ell,t_\ell) 
\right] \rangle
= \left( \det {K}\right)^{1/2}
\int {\em D}[\phi] \exp{ \left[ -{1\over 2} \int \phi({\mathbf q})
 K({\mathbf q},
{\mathbf q}^\prime) \phi({\mathbf q}^\prime) d^3q d^3q^\prime +
i  \sum _{\ell=1}^2 b(t_\ell){\mathbf w}_\ell\cdot \nabla
\phi
\big|_{{\mathbf q_\ell}} \right]}
\label{Eq:path}
\ee
where the kernel $K({\bf q},{\bf q}^\prime)$ represents the functional 
inverse of the two--point correlation function of the field $\phi({\bf q})$.
By defining a six--dimensional vector ${\mathbf c^t}=({\mathbf w}_1, {\mathbf w}_2)$
and choosing the $z$-axis of our reference frame in the direction of the
vector ${\mathbf q}={\mathbf q}_1-{\mathbf q}_2$, we can reduce 
equation (\ref{Eq:path}) to the form:
\be
\langle \exp \left[ -i \sum_{\ell=1}^2 {\mathbf w}_\ell\cdot {\mathbf S}
({\mathbf q}_\ell,t_\ell) 
\right] \rangle
= \exp \left[-{1\over 2} \mathbf c^t M c \right]
\label{Eq:pathsolve}
\ee
where the matrix $\mathbf M$ has the structure
$$
\mathbf {M} =\gamma 
\left( \begin{array} {cccccc}
 b_1^2 & 0 & 0 & b_1 b_2 \psi _\perp & 0 & 0 \nonumber \\
 0 & b_1^2 & 0 & 0 & b_1 b_2 \psi _\perp & 0 \nonumber \\
 0 & 0 & b_1^2 & 0 & 0 & b_1 b_2 \psi _\parallel \nonumber\\
 b_1 b_2 \psi _\perp & 0 & 0 & b_2^2 & 0 & 0  \\
 0 & b_1 b_2 \psi _\perp & 0 & 0 & b_2^2 & 0 \nonumber \\
 0 & 0 & b_1 b_2 \psi _\parallel & 0 & 0 & b_2^2 \nonumber
\end{array}
\right)\eqno (9)
\setcounter{equation}{9}
$$
with $b_i=b(t_i)$ and 
\be
\gamma= {1\over 6 \pi^2} \int _0 ^\infty \!\!\! P(k) dk\;, \ \ \  
\gamma \psi_{\parallel} (q)= {1\over 2 \pi^2} \int _0 ^\infty \!\!\! P(k)
\left[ j_0(kq)-{2\over kq} j_1(kq)\right] dk \;, \ \ \   
\gamma \psi_{\perp} (q)= {1\over 2 \pi^2} \int _0 ^\infty \!\!\! P(k)
{1\over kq} j_1(kq) dk\;,
\ee
having denoted by $j_\ell (x)$
the spherical Bessel function of order $\ell$. 
By substituting this result into equation (\ref{Eq:deldel2}) we can
easily solve the Gaussian integration over the ${\mathbf w}_i$. In order
to perform the remaining integrations, it is convenient to introduce the new
variables ${\bf q}$ and ${\bf Q}={\bf q}_1+{\bf q}_2$. In this way, after some
algebra, we finally obtain: 
\begin {eqnarray}
\lefteqn {1+\xi(r,t_1,t_2) \equiv
1+\langle \delta({\mathbf x}_1, t_1) \delta({\mathbf x}_2,t_2) \rangle=
{1\over  (2\pi )^{1/2}r}
\int _0 ^\infty {q^2 dq\over (b_1 b_2)^{1/2} \gamma
(\psi _\perp -\psi _\parallel)^{1/2} (b_1^2+b_2^2-2 b_1 b_2
 \psi _\perp)^{1/2}}\times} \nonumber\\
& &\ \ \ \ \ \ \ \ \ \ \ \ \ \ \ \ \ \ \ \ \ \ \ \ \ \ \ \  
\times \left\{ D(u_+) \exp\left[ -{(q-r)^2\over 2 \gamma
 (b_1^2+b_2^2-2 b_1 b_2 
\psi _\parallel)}\right] -D(u_-) \exp \left[  
 -{(q+r)^2\over 2 \gamma (b_1^2+b_2^2-2 b_1 b_2 
\psi _\parallel)}\right]
\right\}
\label {Eq:result}
\end{eqnarray}
where $r=|{\mathbf x}_1-{\mathbf x}_2|$, 
\be 
u_{\pm}=\left[ {b_1 b_2 (\psi _\perp -\psi _\parallel) \over
\gamma (b_1^2+b_2^2-2b_1b_2\psi_\perp)(b_1^2+b_2^2-2b_1b_2\psi_\parallel)
}\right] ^{1/2}
\left[{b_1^2+b_2^2-2b_1b_2\psi_\perp\over 2 b_1 b_2 (\psi _\perp -
\psi_ \parallel)}q\pm r \right]
\ee
and $D(x)$ represents the Dawson's integral
\footnote{It is worth stressing that when $\psi_\perp < \psi_\parallel$,
 in order to avoid a complex argument for the Dawson's integral,
it is convenient
to express
the integrand in equation (\ref{Eq:result}) in terms of exponentials and
error functions
(see also the discussion in 
 Schneider \& Bartelmann, 1995).
However, since for the CDM spectrum (the only one considered in our analysis) 
$\psi_\parallel$ is never larger than $ \psi_\perp$, we preferred to write
the solution using $D(x)$.}
(see, e.g., Abramowitz \& Stegun,
1968).
It is straightforward to show that for $t_1=t_2$ the previous formula
reduces to
the usual expression for the mass two--point correlation function in ZA
(Bond \& Couchman 1988, Mann, Heavens \& Peacock 1993, Schneider \&
Bartelmann 1995). 
\begin{figure*}
\vskip -8truecm
\epsfxsize=16cm
%\centerline{\epsfbox{def1.ps}}
\centerline{\epsfbox{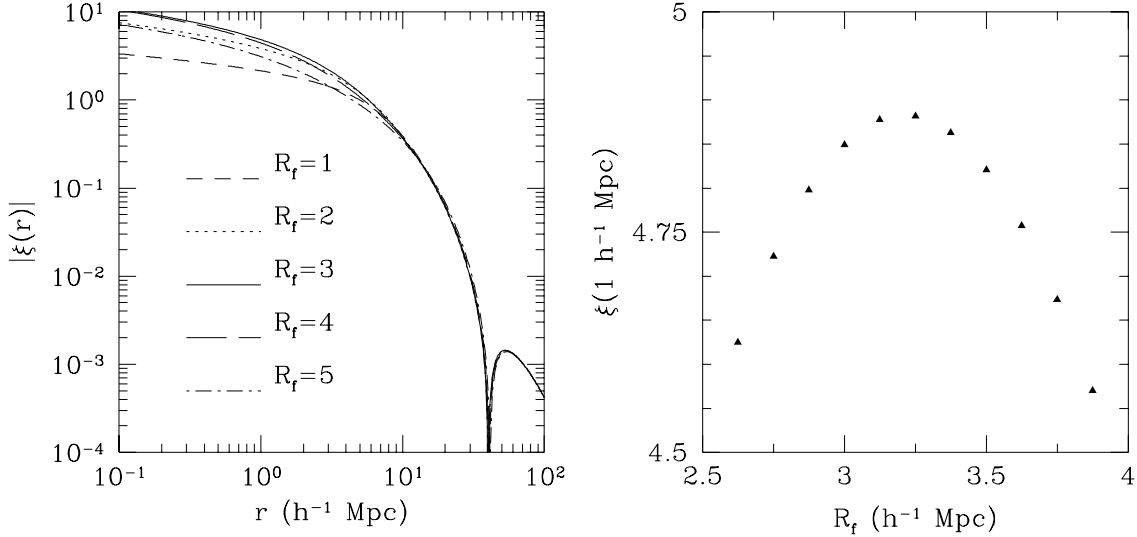}}
\caption[]{
Left panel: the mass autocorrelation
function, obtained using ZA, for a 
{\sl COBE} normalized CDM linear spectrum
is plotted for different values of the truncation
radius $R_f$ (in $\hm$).
Right panel: dependence of the correlation function evaluated at
$r= 1 \hm$ on $R_f$.}
\label{Fig:Rf}
\end{figure*}

We numerically evaluated the two--point correlation function 
$\xi(r,t)\equiv \xi(r,t,t)$
employing a
{\sl COBE} normalized
standard CDM linear power spectrum 
(with density parameter $\Omega=1$ and $h=0.5$).
We used the transfer function of Bardeen et al. (1986) while the
normalization to the four--year {\sl COBE} DMR data is given in Bunn \& White
(1997) and corresponds to $\sigma_8=1.22$.
As already noted by MHP, the small scale behaviour of the resulting correlation
function depends on the value assigned to
the truncation radius, $R_f$,  
defined in equation (\ref{Eq:trunc}) (see Fig. \ref{Fig:Rf}).
If $R_f$ is very small, then shell crossing will not be suppressed
and $\xi(r)$ will show an unusually flat behaviour.
On the contrary, if $R_f$ is too large, the smoothing procedure will
remove an important contribution to the power spectrum, 
causing again too low a correlation. 
Therefore we need a criterion to select $R_f$. Since our main purpose is
to compare
the clustering amplitudes predicted by ZA with those extracted from
the scaling ansatz of JMW, 
we can choose 
$R_f$ so as to optimize the agreement between the
respective correlation functions.
Anyway, we find that this method conforms quite well
to a simpler one already used by MHP: the best $R_f$ is the one that
maximizes $\xi(r,R_f)$ on small scales.
Strictly speaking,
the optimal smoothing radius depends on the scale selected for 
maximizing the correlation: the smaller is $r$ the larger comes out $R_f$
(we find that the difference between the smoothing lengths
obtained by maximizing
$\xi$ at
$r= 0.1 \hm$ and at $ r=1 \hm$ roughly amounts to $0.2 \hm$ and remains
nearly constant by varying $\sigma_8$). However, the effect of this discrepancy
on the correlation evaluated on larger scales is indeed minimal.
Following Schneider \& Bartelmann (1995), we select $r=1 \hm$ as the
scale at which we require $\xi(R_f)$ to be maximal.
As previously stated,
the optimum filtering length increases as the field evolves;
the dependence of the best $R_f$ on $\sigma_8$ 
is almost linear and for $\sigma_8>0.3$
(that in our model corresponds to $z\sim 3$)
it can
be approximated by:
\be
R_f(\sigma_8)=(3.16\, \sigma_8 - 0.65) \hm \;.
\ee
\begin{figure*}
\epsfxsize=8cm
\centerline{\epsfbox{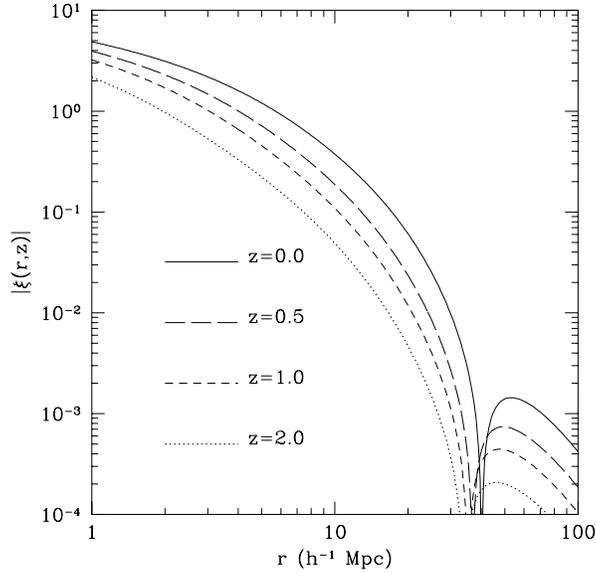}}
\caption[]{
Redshift evolution
of the mass two--point correlation function obtained
using ZA to evolve a linear CDM spectrum.}
\label{Fig:zeldevo}
\end{figure*}

\begin{figure*}
\epsfxsize=8cm
\centerline{\epsfbox{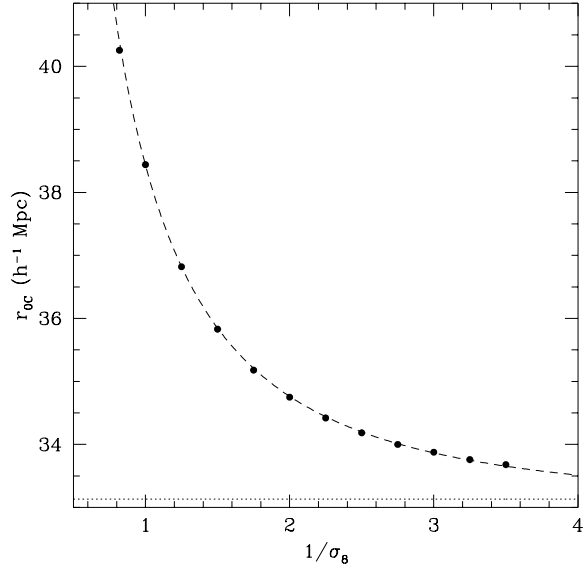}}
\caption[]{
The first zero crossing radius of the correlation function is plotted
against $1/\sigma_8$. The circles represent the results obtained using ZA,
the dashed line
is the fitting function given in the text while the dotted line shows
the prediction of Eulerian linear theory.}
\label{Fig:0cross}
\end{figure*}

The redshift evolution of the correlation
function is shown in Fig. \ref{Fig:zeldevo}.
As expected, on scales that are not affected by shell crossing ($r>R_f$),
$\xi(r,z)$ steepens with decreasing $z$.  
Moreover, we note that the 
first zero crossing radius of $\xi(r,z)$
increases as time goes on (see also Bond \& Couchmann 1988).
A similar pattern has been noticed by Coles (1990) and by Baugh \&
Efstathiou (1994) in the context of second--order Eulerian perturbation theory.
The displacement of the first zero crossing of $\xi$ as a function of time is
plotted in Fig. \ref{Fig:0cross}.
Measuring the degree of dynamical evolution of the density field through
$\sigma_8$, 
this shifting can be described with good approximation
by the function:
\be
r_{\rm 0C}(\sigma _8)-r_{\rm 0C}^{\rm lin} \simeq 
5.3 \, \sigma_8 ^{(1.5+0.1/\sigma_8)} \hm
\ee
where we denoted by $r_{\rm 0C}$ the scale at which the correlation function
crosses for the first time the zero--level and by $r_{\rm 0C}^{\rm lin}$
its linear
counterpart.
It would be interesting to compare this result with the predictions of 
second--order Eulerian (and Lagrangian) perturbation theory and of other
dynamical approximations.

\section {Comparison with the scaling hypothesis}

The analysis of a large set of numerical simulations suggests that, 
in hierarchical models,
the non--linear two--point correlation function, $\xi(r,z)$,
can be related to the linear one, $\xi_{\rm L}(r,z)$, through a simple scaling
 relation
(Hamilton et al. 1991,
Peacock \& Dodds 1994, Jain, Mo \& White 1995,
Peacock \& Dodds 1996).
The main idea is that the action of gravity can be represented
as a continuous change of scale or, better, that the `flow of information'
about clustering propagates along the curves of equation:
\be
r_0=[1+\bar \xi(r,z)]^{1/3} r \;,
\ee 
where $\bar \xi(r,z)$ represents the average correlation function within a
sphere of radius $r$ 
\be 
\bar \xi(r,z) = {3 \over r^3} \int_0^r y^2 \xi(y,z) dy
\ee
and $r_0$ is a sort of Lagrangian coordinate determining a 
`conserved pair surface'
(Hamilton et al. 1991, Nityananda \& Padmanabhan 1994).
In fact, by definition, the average number of 
neighbours of a particle
contained within a spherical volume of radius $r_0$ at the linear stage
(when $\bar \xi \ll 1$) equals the average number
of neighbours  inside a sphere of radius
$r$ in the evolved field.

Here, we want to compare the results obtained in the previous
section, using ZA,  with the predictions of the scaling ansatz 
(hereafter SA) formulated in the
version of JMW:
\be 
\bar \xi(r,z) = B(n_{\rm eff})
 F\left[{\bar \xi_{\rm L}(r_0,z) \over B(n_{\rm eff})}\right]\;
\ee 
with
\be
F(x) = { x + 0.45 x^2 - 0.02 x^5 + 0.05 x^6 \over 1 + 0.02 x^3 + 
0.003 x^{9/2}} \;, \ \ \ \ \ \ 
B(n_{\rm eff}) = \left({3+n_{\rm eff}\over 3}\right)^{0.8}\;,
 \ \ \ \ \ \ 
n_{\rm eff}(z) = \left.{ d \ln P(k)\over d \ln k} \right| _{k_{\rm NL}(z)}         
\ee 
where  $k_{\rm NL}^{-1}$ denotes the radius of the top--hat window function in
which the {\em rms} linear mass fluctuation is unity. 
However, it would be useless to perform the comparison between the
spherically
averaged correlation functions since, on
small scales, $\xi(r)$
obtained using ZA is seriously affected by 
shell crossing 
and the computation of $\bar \xi$ requires an integration starting from $r=0$.
For this reason we prefer to use directly $\xi(r)$. The
two--point correlation function deriving from the ansatz of JMW can
be obtained performing a simple differentiation:
 \be 
\xi(r,z) = { \bigl [1 + B(n_{\rm eff}) F(X) \bigr] F'(X) \Delta \xi_{\rm L}
(r_0,z) 
\over 1 + B(n_{\rm eff}) F(X) - F'(X) \Delta \xi_{\rm L}(r_0,z)} 
+ B(n_{\rm eff}) F(X) \;, \ \ \ \ \ \ X={\bar \xi_{\rm L}(r_0,z)
 \over B(n_{\rm eff})}\; ,
\label{Eq:jmwdiff}
\ee
with $F'(x) = dF/dx$ and 
\be 
\Delta \xi_{\rm L}(r_0,z) \equiv \xi_{\rm L}(r_0,z) - \bar \xi_{\rm L}(r_0,z) 
= {b^2(z) \over 2 \pi^2} \int_0^\infty \!\!\!  k^2 P(k)
\bigl[ j_0(kr_0) - {3 \over kr_0} j_1(kr_0)\bigr] dk \;. 
\ee
\begin{figure*}
\epsfxsize=8cm
\centerline{\epsfbox{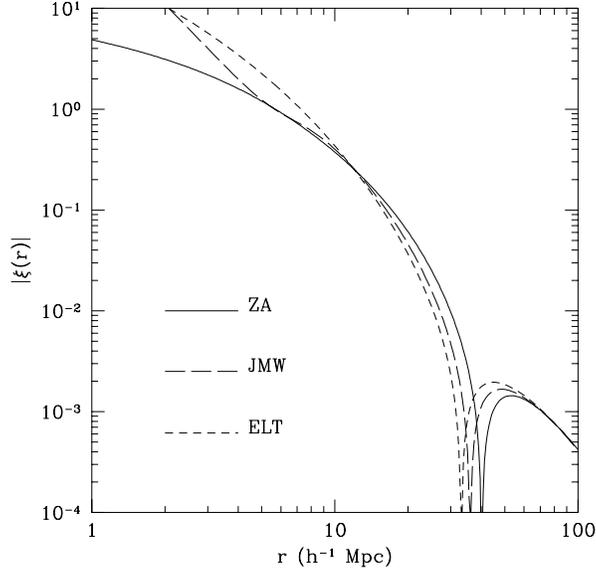}}
\caption[]{
Comparison between the mass autocorrelation functions computed for a CDM
model by using: the Zel'dovich approximation (ZA), the scaling ansatz of
Jain, Mo \& White (JMW) and Eulerian linear theory (ELT).
The linear power spectrum extrapolated to the present epoch ($z=0$) is
normalized to match the {\sl COBE} DMR data ($\sigma_8=1.22$).}
\label{Fig:JMW1}
\end{figure*}
 
\begin{figure*}
\epsfxsize=8cm
\centerline{\epsfbox{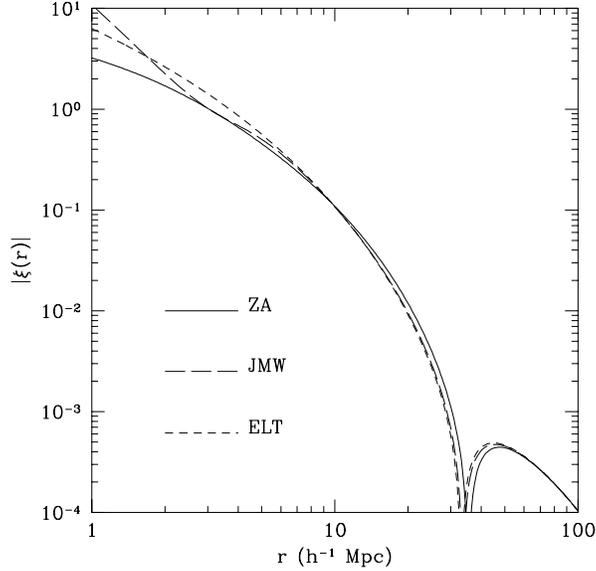}}
\caption[]{
As in Fig. 4, but at $z=1$ ($\sigma_8=0.61$).}
\label{Fig:JMW2}
\end{figure*}

We evaluated the correlation function 
given in equation (\ref{Eq:jmwdiff}) using a {\sl COBE} normalized,
linear CDM spectrum.
In Fig. \ref{Fig:JMW1} we plot the result obtained at $z=0$ with
the corresponding one achieved by using ZA.
For comparison we also show the prediction of Eulerian linear
 theory.
The agreement between ZA and SA
is remarkable on mildly non--linear scales ($4\hm \mincir r \mincir 20 \hm$)
and on completely
linear scales ($r > 50 \hm$).
For example,
 at $r = 5 \hm$, linear theory overestimates
the correlation of JMW by $82 \%$, ZA underestimates it by $2 \%$ while
the accuracy of the JMW fit is about $15-20 \%$.
However, we find that in the interval
$20 \hm \mincir r \mincir 50 \hm$ ZA predicts more non--linear
evolution than SA (for example the $r_{\rm 0C}$ obtained by using ZA
is larger than the one determined through SA).
In order to consider a less evolved field, in Fig. \ref{Fig:JMW2}
we repeat the comparison using the correlation functions evaluated
at $z=1$. Now,
the main item to note is 
that the JMW result 
matches the linear solution
on scales ($r \sim 10 \hm$) 
that, according to ZA, are already involved in non--linear phenomena.

In any case, we do not know the accuracy of the scaling hypothesis 
on large scales. In fact, the function $F(x)$ is obtained by requiring
the resulting $\xi(r)$
to reproduce the linear behaviour where $\bar \xi_{\rm L} \to 0$ and,
simultaneously, to approximate properly the correlation function
extracted from
N--body simulations. However, in order to achieve a detailed description
of non--linear scales, JMW used a relatively small box to perform their
simulations. 
Therefore, imposing the match to linear theory on large scales, 
without having any constraint from numerical data on quasi--linear scales,
could seriously alter
the accuracy of $F(x)$.
 This probably implies that the JMW fitting function could be
improved on large scales.
Our conclusion is shared by
Baugh \& Gazta\~{n}aga (1996, hereafter BG), who tested the scaling ansatz
for the evolution of the power spectrum
 against the results
of 5 N--body simulations performed within a  $378 \hm$ box.
Indeed, they found
that the JMW formula gives a relatively poor description
of the large--scale behaviour
even though the agreement between the spectra 
remains always within the quoted 20 \% accuracy.

\begin{figure*}
\epsfxsize=8cm
%\centerline{\epsfbox{bau.ps}}
\centerline{\epsfbox{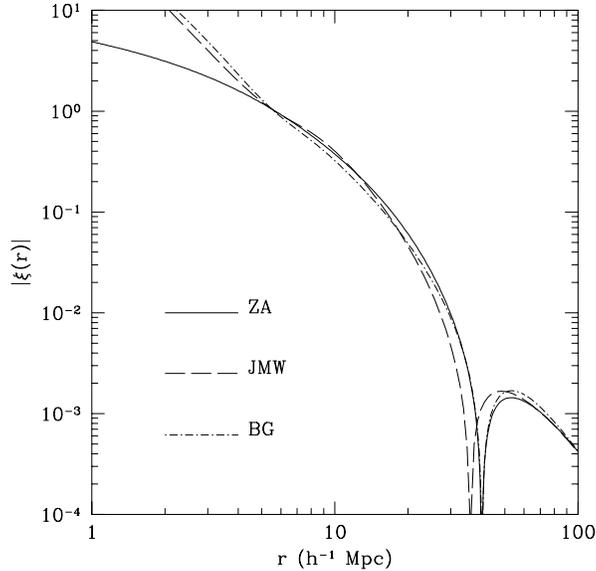}}
\caption[]{
Mass two--point correlation functions at the epoch in which $\sigma_8=1$
obtained from a linear CDM spectrum evolved through the Zel'dovich
approximation (ZA) and through the scaling ans\"atze by Jain, Mo
\& White (JMW) and by Baugh \& Gazta\~naga (BG).}
\label{Fig:BAU}
\end{figure*}

By using the output of their simulations,
BG proposed a new scaling formula calibrated on large scales.
As initially suggested by Peacock \& Dodds (1994), the analytic
expression of this SA concerns the dimensionless power spectrum, 
$\Delta^2(k,z)=k^3 P(k,z)/ 2 \pi^2$
(i.e. the contribution to the the variance of the density contrast 
per bin of $\ln k$),
 while,
following JMW, it takes account of a spectral dependence of the 
transformation:
\be \Delta^2(k,z)= \beta(n_{\rm eff}) f\left[ 
{\Delta^2_{\rm L}(k_{\rm L},z)\over \beta(n_{\rm eff})} \right]\;,
\ \ \ \ \ \ 
k_{\rm L}=\left[ 1+\Delta^2(k,z)\right]^{-1/3} k
 \ee
where
\be 
f(x)=x \left( {1+0.598x-2.39x^2+8.36x^3-9.01x^{3.5}+2.895x^4
\over 1-0.424x+[2.895/(11.68)^2]x^3}\right)^{1/2}\;, \ \ \ \ \ \ 
\beta(n_{\rm eff})=1.16\left( {3+n_{\rm eff}\over 3}\right) ^{1/2} \ee
and the subscript {\rm L} marks linear quantities.
The function $f(x)$ has been obtained by matching the power spectrum in
the simulations at $\sigma_8=1$,
with an accuracy of $5 \%$, over the range $0.02 \, h\,{\rm Mpc}^{-1}< k <
 1.0 \, h \,
{\rm Mpc}^{-1}$ and by forcing the fit to have the asymptotic form
$f(x) \to 11.68 \,x^{3/2}$ when $x \to \infty$ (Hamilton {\it et al.}
 1991).
The two--point correlation function is related to
$\Delta^2(k,z)$ through
the Fourier relation:
\be \xi(r,z)=\int_0^\infty  \Delta^2(k,z) j_0(kr)
 {dk\over k}.
\ee
In Fig. \ref{Fig:BAU} we compare the correlations obtained by
using ZA and the JMW formula with the results of the scaling ansatz by BG:
we are considering a standard CDM linear spectrum at the epoch in which
$\sigma_8=1.22$.
We immediately note that using larger simulation boxes to calibrate the SA
allows a better
determination of the correlation function for
$r \magcir 20 \hm$.
%In fact we find that the correlations obtained with ZA and with the BG formula
%agree by better than $ 20 \%$
%over the ranges  
%$4 \hm \mincir r \mincir 36 \hm$ and $ r \magcir 50.5 \hm$ 
%while the discrepancy between ZA and the JMW ansatz is less
%than $ 20 \%$ for
%$3.5 \hm \mincir r \mincir 17.5 \hm $ and $ r \magcir 47.5 \hm$.
In fact, we find that the correlations obtained with ZA and with the BG formula
agree by better than $ 20 \%$ for $r > 4.6 \hm$ (with
the exception of a very small $r$-interval centred in the first zero crossing
of $\xi$) while
the discrepancy between ZA and the JMW ansatz is less
than $ 20 \%$ over the ranges
$4.1 \hm < r < 18.3 \hm $ and $ r > 49.6 \hm$.
Similar patterns are obtained considering different values of $\sigma_8$.
This shows that the BG fit, that has been calibrated against
large box CDM simulations, gives also a very good description 
of the mass clustering predicted by ZA on intermediate scales.
In any case, as expected, the JMW formula is sensibly more accurate for
$5 \hm \mincir r \mincir 15 \hm$ where the BG predictions
grow worse as $\sigma_8$ assumes values significantly larger than 1.

On the other hand, it would be interesting
to check the reliability of ZA and second--order Eulerian perturbation
theory by directly comparing their predictions on these scales.
Bond \& Couchmann (1988), studying the weakly non--linear evolution of
the CDM
power spectrum, found
remarkable agreement between the two approximations.
Moreover, Baugh \& Efstathiou (1994) 
showed that second--order Eulerian
perturbation theory can reproduce, at least qualitatively,
the evolution of the power spectrum predicted by numerical simulations.
However,
Jain \& Bertschinger (1994)
found that the agreement between perturbation theory and N--body outcomes
gets worse as the density field evolves. Besides, their results 
are inconsistent with the low--$k$ behaviour
of the second--order Eulerian correction to the
CDM power spectrum computed by Bond \& Couchmann (1988),
raising again the issue about the compatibility between ZA and
perturbation theory. 
In a recent work concerning the evolution of scale invariant spectra,
Scoccimarro \& Friemann (1996b) showed that, if the spectral index $n$ 
satisfies $-3<n<-1$,
Eulerian perturbation theory is able to reproduce fairly well the
power spectrum obtained though the scaling ansatz, while the one--loop
perturbative version of ZA gives worse results.
Anyway, Bharadwaj (1996a,b) pointed out that the effects of multistreaming
on the correlation function 
cannot be studied perturbatively. 
This fact implies that our result, obtained considering the full Zel'dovich
approximation,
should be more reliable than any other achieved by adopting a
perturbative version of ZA.
In any case,
it would be interesting to clarify to which extent ZA and
Eulerian perturbation theory agree on large scales.

\section {The correlation of high redshift objects}

In this section, we study the evolution of the
cross correlation function of the mass density contrast evaluated
at two different times as defined in equation (\ref{Eq:result}).
This quantity could play an important role in comparing
the clustering properties extracted from deep redshift surveys
to the predictions of theoretical models for structure formation.
In practice, one always collects data on correlations in a finite
redshift strip of his past light cone
while the quantity $\xi(r,t)$, normally used in theoretical works, refers
to objects
selected on an hypersurface of constant cosmic time.
Therefore, as far as one is considering a deep sample of cosmic objects,
it is not correct to relate the observed clustering properties to
$\xi(r,t)$. This issue is addressed in detail
by Matarrese et al. (1997, hereafter MCLM)
who build
a theoretical quantity that allows a direct comparison
of model predictions to the observed correlations.
Their approach can be
divided into three steps:
first of all they compute the redshift evolution of mass correlations,
then they relate the clustering properties of cosmic objects to
the matter distribution by means of a linear bias relationship
and finally they convolve the result with the observed
redshift distribution of the class of objects under analysis.
By assuming that the effects of redshift distortions and of the
magnification bias due to weak gravitational lensing are negligible and 
by considering isotropic selection functions,
MCLM showed that
the theoretical estimate for the observed two--point correlation
function can be formally expressed as an integral over 
$z_1$ and $z_2$  
of the 
function $\xi(r,z_1,z_2)$ weighted by geometrical factors and effective
bias parameters (all dependent on $z_1$ and $z_2$).
Different classes of objects are selected by changing the
amplitude and the redshift dependence of the effective bias.
However, in the absence of a model for the evolution of 
the cross--correlation,
only assuming that the above mentioned integral is dominated by
the contribution of 
objects whose redshifts are nearly the same,
can one
estimate the observed correlation function deriving from
a particular scenario of structure formation.
In this way, one is
allowed to replace
$\xi (r,z_1,z_2)$ with
$\xi(r,\bar z)$, where 
$\bar z$ is a suitably defined average between $z_1$ and $z_2$ 
that, for simplicity, MCLM
identify with
$\bar z=(z_1+z_2)/2$.
This is a crucial approximation, as it allows MCLM
to use the JMW ansatz to compute the non linear mass correlation
function
(there is no known scaling ansatz for $\xi(r,z_1,z_2)$).
However, as shown in the previous paragraphs, ZA allows the computation
of $\xi(r,z_1,z_2)$ so that
we are able to compute the theoretical estimate for the observed 
correlation function by using both the complete and the approximated formulae
given by MCLM (respectively their equations 15 and 18).
Therefore we can check here, within the validity of ZA, the reliability of the
approximation introduced by MCLM.
Large discrepancies between the exact and the approximated correlations
would obviously invalidate their whole analysis and consequently
also their complete formula for $\xi_{\rm obs}$ would be unutilizable.
On the other hand, if the approximated correlation function turns out
to reproduce accurately the complete one, MCLM formulae could represent
an important tool to disprove cosmological models
in the light of present and future observations.

\begin{figure*}
\epsfxsize=8cm
\centerline{\epsfbox{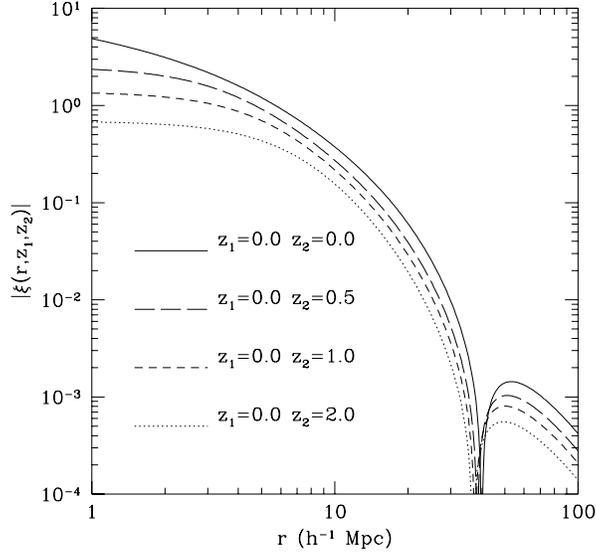}}
\caption[]{
Cross correlation between the density contrast field
evaluated at two different redshifts vs. comoving separation.}
\label{Fig:duet}
\end{figure*}

In order to compute $\xi(r,z_1,z_2)$ using equation (\ref{Eq:result}), we 
truncated the linearly extrapolated power spectrum 
$b(z_1) b(z_2) P(k)$
according to the prescription:
\be
P_{T}(k,z_1,z_2)=b(z_1) b(z_2) P(k)\exp{\left[ -k^2 R_{f}(z_1) R_{f}(z_2)\right]}
\label{Eq:t2D}
\ee
where $R_{f}(z)$ represents the optimum filtering length for the
density field at redshift $z$, determined by following the method described
in Section 3.
On small scales,
the correlation functions that we obtain
 opting for this truncation procedure appear
much more
flattened than those computed at a single time. 
The evolution of $\xi(r,z_1,z_2)$ as $z_2$ changes is shown, for a CDM model, in
Fig. \ref{Fig:duet}.
It is evident that even though the correlation decreases
as $z_2$ grows, its decay is very slow.
Actually, the ratios between the correlations computed at the same $r$, 
for different
pairs of redshifts, are very similar to the predictions of linear theory.
We find that the redshift evolution of the cross correlation function can
be approximately described by the relation:
\be
 \xi(s,z_1,z_2) \simeq \left[ \xi(s,z_1) \xi(s,z_2)\right]^{1/2} 
\left[ 1-2\Theta(s-1) \right]
\label {Eq:approx2t}
\ee
where the quantity $s=r/r_{\rm 0C}(z)$ is introduced in order to
take into account the shifting of the first zero crossing
of $\xi(r,z)$
and $\Theta(x)$ is the Heaviside step function.
Moreover, the first zero crossing radius of $\xi(r,z_1,z_2)$
is nearly given by the geometric average of
$r_{\rm 0C}(z_1)$ and $r_{\rm 0C}(z_2)$.
For $s>0.1$   
equation (\ref{Eq:approx2t}),
which is meaningful up to the scale at which the first of the two $\xi(s,z)$
reaches its second zero crossing,
 reproduces $\xi(s,z_1,z_2)$ with an accuracy of 
$\sim 5 \%$.
Anyway, for $s \magcir 2$, the usual relation 
$ \xi(r,z_1,z_2) \simeq [\xi(r,z_1) \xi(r,z_2)]^{1/2}{\rm sign}[\xi(r,z_1)]$
deriving from linear
theory is preferable.

We can now check the accuracy of the approximation introduced by MCLM 
that consists in computing the
theoretical estimate for the
observed correlation function
by replacing $\xi(r,z_1,z_2)$ with $\xi(r,\bar z)$, where $\bar z=(z_1+z_2)/2$,
in the appropriate formula.
For simplicity
(and in order to isolate the phenomenon of clustering evolution) 
we will assume no bias, no selection effects and a constant comoving
number density in an Einstein--de Sitter universe.
In this case, equation 15 of MCLM reduces to:
\be
\xi_{\rm obs}(r,z_{\rm min},z_{\rm max})=
{\displaystyle \int_{z_{\rm min}}^{z_{\rm max}} 
{2+z_1-2(1+z_1)^{1/2} \over (1+z_1)^{5/2}} \,
{2+z_2-2(1+z_2)^{1/2} \over (1+z_2)^{5/2}} \,
\xi(r,z_1,z_2) \,  dz_1 dz_2 \over
\displaystyle \left[ \int_{z_{\rm min}}^{z_{\rm max}} 
{2+z-2(1+z)^{1/2} \over (1+z)^{5/2}}\, dz
\right] ^2}
\label{Eq:mat}
\ee
where we denoted by $\xi_{\rm obs}(r,z_{\rm min},z_{\rm max})$ the
(ensemble averaged) theoretical estimate for the two--point
correlation function measured by an observer that acquires data from
the region of his past light cone corresponding to
the redshift interval $ [z_{\rm min},z_{\rm max}]$.

Considering only the linear evolution of density fluctuations,
$\xi(r,z_1,z_2)=\xi(r,0,0)/[(1+z_1)(1+z_2)]$,
the integrals contained in equation (\ref{Eq:mat}) can
be analytically performed.
In this case, the quantity $\xi_{\rm obs}(r,z_1,z_2)/\xi(r,0,0)$ does not
depend on $r$; for example we obtain  
$\xi_{\rm obs}(r,0,2)/\xi(r,0,0)\simeq 0.224$ and
$\xi_{\rm obs}(r,0,1)/\xi(r,0,0)\simeq 0.375$.
In this regime, we find that the approximation for $\xi_{\rm obs}$ introduced
by MCLM is accurate to $2-3 \%$.

\begin{figure*}
\epsfxsize=8cm
\centerline{\epsfbox{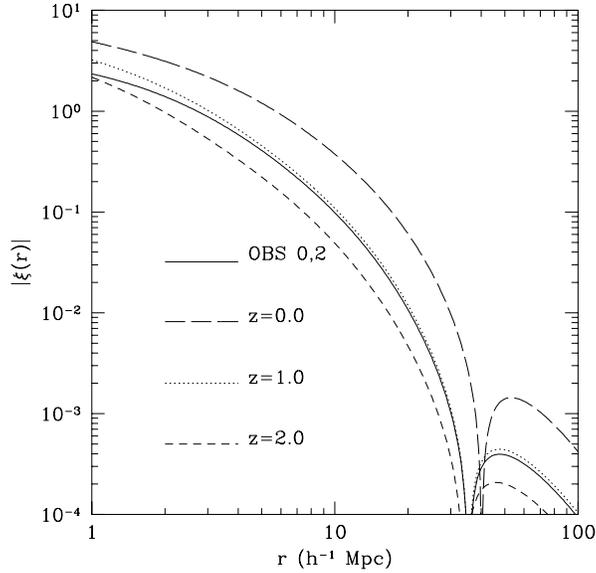}}
\caption[]{
The observed two--point correlation function computed using equation (26)
for a CDM
model with
$[z_{\rm min},z_{\rm max}]=[0,2]$. For comparison, the corresponding $\xi(r,z)$
evaluated for $z=0,1,2$
are plotted.}
\label{Fig:integrata}
\end{figure*}

In order to extend our analysis also to the mildly non--linear evolution,
we numerically computed $\xi_{\rm obs}$ by using the cross correlation
given in equation (\ref{Eq:result}). The result obtained 
for $ [z_{\rm min},z_{\rm max}]=[0,2]$
is shown in Fig. \ref{Fig:integrata}: also in this case
$\xi_{\rm obs}$
looks like the usual correlation function  evaluated at some
intermediate redshift.
We then tested the accuracy of the above mentioned 
simplified scheme for the computation of
  $\xi_{\rm obs}$,
finding good agreement between the exact and the rough estimates
(excluding a small neighbourhood of the
zero--crossing radius of $\xi(r,z_1,z_2)$,
 where the approximated method breaks down,
we find a maximum discrepancy of $6 \%$
for  $[z_{\rm min},z_{\rm max}]=[0,2]$ and of $3 \%$ for  
$[z_{\rm min},z_{\rm max}]=[0,1]$).
Anyway, the simplified procedure to compute $\xi_{\rm obs}$
can be further improved:
adopting
a different way of performing the average between redshifts, namely
$1+\bar z=[(1+z_1)(1+z_2)]^{1/2}$, ensures more accurate predictions 
(in this case the maximum error is always of the order of $1\%$).
Probably this higher precision is due to the fact that we are considering mildly non--linear
scales and the latter approximation gives exact results for linear
evolution.

\section{Summary}

In this paper,
we have studied in detail the evolution of the mass two--point correlation
function 
by describing the growth of density perturbations through ZA.
Our motivations were originated by the well known ability
of ZA to reproduce the weakly non--linear regime of 
gravitational dynamics.
On scales that are not affected by shell--crossing, we found that
the correlation function steepens as the clustering amplitude
increases. Moreover,
we showed that non--linear interactions are able to move the first
zero crossing
of $\xi(r)$ and we gave a quantitative description of this
shifting for a CDM linear spectrum.

We then compared our results with the predictions of the scaling ansatz
for clustering evolution formulated by JMW, obtaining remarkable
agreement between the correlations on mildly non--linear scales
and on completely linear scales.
However, between these two regimes, the
JMW prescription, which has been obtained requiring the resulting correlation
to reproduce the linear behaviour on large scales, predicts
smaller clustering amplitudes than ZA. 
We think that this disagreement is caused by the smallness of the box
used by JMW to perform their N--body simulations.
Actually, imposing to match the linear solution where
$\bar \xi_{\rm L} \to 0$, without
having any constraint from numerical data on quasi--linear scales, could alter
the accuracy of the fitting function that embodies the scaling ansatz.
In connection with this hypothesis, we compared ZA predictions on 
correlations with the output of a different scaling ansatz calibrated
against large box simulations by BG.
In effect, on large scales, the BG formula agrees better with
ZA, keeping the same accuracy of the JMW fit on intermediate scales.

On the other hand, the reliability of ZA on these scales and for dynamically
evolved fields ($\sigma_8 \magcir 1$) should be verified by directly comparing
its predictions with the results of other approximations and numerical 
simulations.

Finally,
we studied the evolution of the cross correlation between the density
field evaluated at two different epochs and,
adopting the method introduced by MCLM,
we used our results to
compute the theoretical prediction for the observed correlation function
deriving from a deep catalogue of objects.
In this context,
we proposed a simplified procedure for 
the computation of $\xi_{\rm obs}$ that, at least for 
quasi--linear scales, significantly improves another approximation
previously introduced by MCLM.
This result confirms that the MCLM method can be used to make
quantitative predictions about clustering evolution that find
a direct observative counterpart in the analysis of deep surveys.

\section*{Acknowledgments.}
I would like to thank Sabino Matarrese for the encouragement and the useful
suggestions.
I am grateful to the referee, Carlton Baugh, for helpful comments on
the manuscript.
Francesco Lucchin, Lauro Moscardini and Pierluigi Monaco are also thanked
for discussions.
Italian MURST is acknowledged for financial support. 

%\clearpage

%\newpage

\end{document}